# Dependence of Substrate Irradiation Reaction Rate Stimulation on Lactic Dehydrogenase Source

George E. Bass* and James E. Chenevey
5567 Ackerman Cove, Bartlett, TN USA

*Author to whom correspondence should be addressed.  Email: gbass@utmem.edu

**Abstract**

Stimulation of LDH initial reaction rates by timed pre-irradiation of crystalline sodium pyruvate and lithium lactate is reported for enzymes isolated from rabbit muscle, pig heart, human erythrocytes and chicken heart.  The phenomenon investigated is referred to as the Comorosan effect.  For the mammalian source enzymes, the pyruvate irradiation stimulations occurred at irradiation times of 5 and 35 sec. and the lactate irradiation times at 15 and 45 seconds.  In contrast, for the chicken heart enzyme, the pyruvate irradiation stimulations occurred at 15 and 35 sec., while those for lactate occurred at 5 and 20 sec.  Thus, a shift in stimulatory irradiation times is found on going from the mammalian enzymes to the avian enzyme.  A similar shift between mammalian and yeast enzymes has been established by Comorosan and co-workers.  For the chicken heart LDH, the separation between successive irradiation times is different for the forward and reverse reactions.  This is the first reported incidence of the separation not being the same.

**Introduction**

Comorosan and co-workers (Comorosan et al.1970a,b,c; 1971a,b) have reported experimental studies in which the rate of in vitro enzymic reactions could be enhanced by irradiating the enzyme substrate in the crystalline state prior to dissolution and incorporation in the reaction mixture.  The reaction rate increases occurred only for certain timed exposures of the pure crystalline materials.  Initially, an x-ray source was employed.  This was soon replaced by a high pressure mercury lamp.  With increasing precision in their kinetic assays, it became apparent that the special irradiation times, designated t*, that produced increases for a given enzyme – substrate reaction rate could be represented by the manifold  $t^* = t_m + n\tau$ where $t_m$  is the shortest effective exposure time, $\tau$ is a constant and n is an integer (0, 1, 2, …).  All t* have been found to be a multiple of 5 sec.  Results from over 50 enzyme – substrate reactions involving enzymes isolated from 6 different species have been reported (Comorosan et al., 1980).

Of particular interest here are the relative values of t* found for forward and reverse reactions catalyzed by a given enzyme.  For example, the enzyme lactic dehydrogenase will catalyze the conversion of pyruvate to lactate as well as the conversion of lactate to pyruvate.  In all of the studies reported to date, t* values which stimulate the reaction in one direction have no effect on the reaction in the reverse direction.  The $t_m$ values for the forward and reverse reactions have always been found to be different while the **τ** values have always been found to be the same for a given enzyme isolate. Comorosan termed this characteristic a "flip-flop" behavior and pointed out that it might reflect an unrecognized energy-mediated biological control mechanism (Comorosan, 1970c). Existing examples of this behavior are presented in Table 1. Systems of related enzymes were found to display patterns in the $t_m$, τ values consistent with producing metabolic flux through the system (Comorosan et al.,1971a,b). That is, all of the reactions leading to the primary product of the system tend to be stimulated by a common energy-mediated "signal," while reactions that served to divert intermediate metabolites elsewhere tend to respond to different signals.

It has been found that the t* values for a particular enzyme can vary with the source (biological species) of the enzyme. Most notably, Comorosan and coworkers found completely different sets of t* values for corresponding yeast and rabbit muscle enzymes comprising the glycolysis, gluconeogenesis and citric acid cycle pathways (Comorosan et al.,1971a,b).

The studies reported here were undertaken to extend characterization of the LDH forward and reverse reactions to additional species.

**Materials and Methods**

The Lactate – LDH Reaction
Identical 15 mg samples of lithium lactate (Sigma Chemical Co.) were prepared by pipetting 0.25 ml of a 50 mg/ml aqueous solution into identical small containers (Falcon 3001, 35x10 mm tissue culture dishes) which were placed in a vacuum desiccator at room temperature until dry (at least 6 h). These samples were irradiated for specific times with a narrow band selected from the output of a high pressure mercury lamp (General Electric H 100 A4/T, $\lambda_{max}$ = 546.1 nm, bandwidth = 8.8 nm, Detric Optics 2-cavity bandpass filter). The distance of the lamp above the sample (approximately 19 cm) had been adjusted to produce an illuminance of 400 – 500 footcandles (Panlux Electronic Footcandle Meter) at the sample. The irradiated (and non-irradiated control) samples were dissolved in one ml of distilled water (3 min allowed for dissolution) and 0.1 ml aliquots placed in 1 cm square cuvettes along with 2.9 ml of 0.1 mg/ml NAD (Sigma Chemical Co.) / 0.05 M pH 7.5 phosphate buffer. Reaction was initiated in this solution, in place in the spectrophotometer, by addition of 0.050 ml of approximately 20 units/ml LDH (Sigma Chemical Co., Rabbit Muscle Type II, Rabbit Muscle Type V (M4 isozyme), Beef Heart, Porcine Heart, and Chicken Heart). Absorbance change at 340 nm for the first 12 sec of reaction, $\Delta A_{340}$, was determined from the continuous, approximately linear absorbance recording (Gilford 2400S, chart speed 0.1 min/in., 0.1 absorbance units full scale). All experiments were conducted at room temperature (approximately 23 C). In order to take into consideration any variations in room temperature and to offset possible fluctuations in enzyme and NAD activities over the course of a day's experiments (and from one enzyme solution batch to another) the lithium lactate samples were assayed in groups of four, at least one of which was always a non-irradiated control. In the following, each of these groups of four samples is referred to as a "Run." In every Run, assay of the sample(s) irradiated for a t* time was always preceded and followed by assay of samples which were not irradiated or which were irradiated for non-t* times (also considered "control" samples). This allows distinction between irradiation-induced activation and possible activity drift of one sort or another in the course of a Run. In a given Run, all samples were first irradiated for the times indicated, then all were dissolved and loaded into cuvettes as outlined above. Following addition of NAD/buffer solution to all cuvettes, the four samples were assayed enzymatically within a lapse time of 10 min. The rapid manual addition and mixing of the enzyme was accomplished in 1.5 to 2.0 sec. Lapse time from beginning of the crystalline lactate dissolution step to completion of the enzyme assays was always less than 30 min. Dissolution of the samples was usually initiated within 5 – 10 min. after the irradiation step, always within one hour. The crystalline lithium lactate samples were always used within 36 h following their preparation.

The Pyruvate – LDH Reaction
Experiments with sodium pyruvate (Sigma Chemical Co.) were conducted in a manner entirely parallel to the lithium lactate assays. Three mg sodium pyruvate samples, 0.1 mg/ml NADH (Sigma Chemical Co.) / 0,050 M pH 7.5 phosphate buffer, and 0.050 ml of approximately 40 units/ml LDH were employed as described above for the lactate reaction.

**Results**
The Lactate – LDH Reaction

The lactate – LDH reaction of four mammalian LDH enzymes, rabbit muscle $M_4$ isozyme, human RBC, beef heart and porcine heart, was investigated. The $t_m$, $\tau$ values for irradiated crystalline lithium lactate were found to be identical, $t_m = 15$ sec and $\tau = 30$ sec, for these enzyme sources. Results are provided in Table 2 for irradiation times of 0, 14, 15, 16, 44, 45 and 46 sec. (In Tables 2, 3 and 5, the irradiation times are represented as 0, t*-1, t*, and t*+1 for each "Run" which consisted of the analysis of four crystalline samples. Three replicate Runs are presented for each t*.)

The lactate – chicken heart LDH reaction responded differently from the mammalian enzymes. Here, the reaction rate was found to be enhanced by lithium lactate irradiations having $t_m = 5$ sec and $\tau = 15$ sec. Results for irradiation times of 0, 4, 5, 6, 19, 20, and 21 sec are shown in Table 3. These results are distinctly different from those found for LDH isolated from mammalian tissues.

In order to more directly confirm the divergence of the chicken LDH $t_m$, $\tau$ values from those found for mammalian LDH sources, a series of experiments were conducted wherein chicken and rabbit muscle type II LDH were both utilized in the same Run. Each Run involved measurement of the absorbance change in 12 sec reaction time for four lithium lactate samples, two irradiated and two not. Two samples, one irradiated and one not, were assayed with rabbit enzyme and two with chicken enzyme. The order in which the two enzymes were assayed in a Run was varied (rabbit first, chicken second, and conversely) to guard against possible sequencing bias. Results are summarized in Table 4. As before, the chicken lactate – LDH reaction was found to be stimulated by lithium lactate irradiations of 5, 20 and 35 seconds, corresponding to $t_m$ = 5 sec, $\tau$ = 15 sec. The rabbit LDH was not stimulated for these irradiation times, but was for irradiation of 15 sec., in keeping with its previously reported parameters of $t_m$ = 15 sec, $\tau$ = 30 sec .

The Pyruvate – LDH Reaction

The pyruvate – LDH reaction of two mammalian LDH enzymes, rabbit muscle $M_4$ isozyme and human RBC was investigated. The $t_m$, $\tau$ values for irradiated crystalline sodium pyruvate were found to be identical, $t_m$ = 5 sec and $\tau$ = 30 sec, for these enzyme sources. Results are provided in Table 5 for irradiation times of 0, 4, 5, 6, 34, 35 and 36 sec.

The pyruvate – chicken heart LDH reaction responded differently from the mammalian enzymes. Here, the reaction rate was found to be enhanced by sodium pyruvate irradiations having $t_m$ = 15 sec and $\tau$ = 20 sec. Results for irradiation times of 0, 5, 10, 15, 20, and 25 sec are shown in Table 6. These results are distinctly different from those found for LDH isolated from mammalian tissues.

In order to more directly confirm the divergence of the chicken LDH $t_m$, $\tau$ values from those found for mammalian LDH sources, a series of experiments were conducted wherein chicken and rabbit muscle type II LDH were both utilized in the same Run. Each Run involved measurement of the absorbance change in 12 sec reaction time for four sodium pyruvate samples, two irradiated and two not. Two samples, one irradiated and one not, were assayed with rabbit enzyme and two with chicken enzyme. The order in which the two enzymes were assayed in a Run was varied (rabbit first, chicken second, and conversely) to guard against possible sequencing bias. Results are summarized in Table 7. The chicken pyruvate – LDH reaction was found to be stimulated by sodium pyruvate irradiations of 15, 35 and 55 sec, corresponding to $t_m$ = 15 sec, $\tau$ = 20 sec. The rabbit muscle LDH was stimulated for irradiation times of 5 and 35 sec, but not for irradiations of 15 and 55sec, in keeping with its previously reported parameters of $t_m$ = 5 sec, $\tau$ = 30 sec.

**Discussion**

The reaction rate stimulation parameters, $t_m$ and $\tau$, for rabbit muscle LDH Type II (Sigma Chemical co.) found here for the forward and reverse reactions are the same as previously reported from this laboratory (Bass, G.E. and Chenevey, J.E., 1977). The $t_m$ value found here for the lactate reaction ($t_m$ = 15 sec) differs from that reported by the Comorosan group (Comorosan, et al., 1971b) for the rabbit muscle enzyme isolated in their laboratory ($t_m$ = 20 sec). The reason for this difference has not been determined. Identical $\tau$ values (30 sec) were found by both laboratories for the forward and reverse reactions. Both laboratories found $t_m$ = 5 sec for the pyruvate reaction.

The finding here of different values for the $\tau$ parameter for the forward and reverse reactions with chicken heart LDH ($\tau$ = 15 sec for lactate, $\tau$ = 20 sec for pyruvate) is the first noted deviation from Comorosan's "flip-flop" behavior. Thus, both the pyruvate and the lactate reactions are stimulated for $t^*$ = 35 sec. This is notable because of its implications for any future efforts to derive a code relating $t_m$, $\tau$ parameters to an *in vivo* biological control process.

In this study, identical values of $t_m$ = 5 sec, $\tau$ = 30 sec were found for the pyruvate reaction catalyzed by all five of the mammalian source enzymes studied. Similarly, $t_m$ = 15 sec, $\tau$ = 30 sec was found for the lactate reaction catalyzed by the mammalian enzymes studied in this laboratory. Different values were found, however, for chicken LDH. Thus, in this case, a ($t_m$, $\tau$) parameter set shift was found at the "class" level within the animal kingdom. Comorosan et al. previously found such a shift at the kingdom level between Fungi (yeast) and Animalia (rabbit). The irradiation reaction rate stimulation phenomenon has been demonstrated in enzymes isolated from species belonging to the Fungi, Animalia, and Plantae (jack bean urease) kingdoms of the Eucarya Domain and bacteria (penicillinase, B. subtilis) of the Eubacteria Domain (Sherman et al., 1974). Irradiated crystalline growth factors and antibiotics have been found to stimulate and inhibit growth of yeast and bacteria in a similar manner (Comorosan et al., 1970, 1973; Bass and Crisan, 1973; Sherman et al., 1974). We are thus led to observe that this phenomenon appears, at this early juncture, to involve a species-dependent characteristic of enzymes that is compatible with biological evolution across most of its spectrum (the Archaea Domain has not been investigated).

Table 1. Previously Reported $t_m$ and $\tau$ Parameters for Forward and Reverse Enzyme Reactions

| Enzyme | Source | Irradiated Substrate | $t_m$ sec | $\tau$ sec |
|---|---|---|---|---|
| Lactic dehydrogenase | Rabbit muscle | Pyruvate[a,b,c,d] | 5 | 30 |
| | | Lactate[a] | 20 | 30 |
| | | Lactate[b] | 15 | 30 |
| | Rat liver[e] | Pyruvate | 5 | 30 |
| | | Lactate | 15 | 30 |
| | Yeast[f] | Pyruvate | 5 | 20 |
| | | Lactate | 10 | 20 |
| | | | | |

| Malic dehydrogenase | Rabbit muscle[a] | Malate | 10 | 35 |
|---|---|---|---|---|
| | | Oxaloacetate | 15 | 35 |
| | Rat liver[e] | Malate | 10 | 35 |
| | | Oxaloacetate | 15 | 35 |
| | Yeast[f] | Malate | 5 | 20 |
| | | Oxaloacetate | 10 | 20 |
| | | | | |
| Malic enzyme | Rabbit muscle[a] | Pyruvate | 15 | 30 |
| | | Malate | 20 | 30 |
| | Yeast[f] | Pyruvate | 5 | 20 |
| | | Malate | 10 | 20 |
| | | | | |
| Gutamate pyruvate transaminase | Rabbit muscle[a] | Alanine | 15 | 30 |
| | | Pyruvate | 20 | 30 |
| | | | | |
| Phosphoglucomutase | Rabbit muscle[a] | Glucose-6-phosphate | 15 | 30 |
| | | Glucose-1-phosphate | 20 | 30 |

[a]Comorosan et al., 1971b
[b]Bass and Chenevey, 1977
[c]Sherman et al., 1973
[d]Goodwin and Vieru, 1975
[e]Comorosan et al., 1980
[f]Comorosan et al., 1971a

Table 2. Lactate – Mammalian LDH Reaction $t_m$, $\tau$ Parameters

| | | | ΔA340 / 12-sec enzyme reaction | | | | |
|---|---|---|---|---|---|---|---|
| **Source** | **Run** | **t*** | **0 sec** | **t*-1 sec** | **t* sec** | **t*+1 sec** | **Δ(ΔA)** |
| Rabbit Muscle Type V (M$_4$) | 1 | 15 sec | 0.0601 | 0.0600 | 0.0615 | 0.0602 | +0.0014 |
| | 2 | | 0.0605 | 0.0606 | 0.0628 | 0.0608 | +0.0022 |
| | 3 | | 0.0590 | 0.0586 | 0.0605 | 0.0588 | +0.0017 |
| | | | | | | | |
| | 4 | 45 sec | 0.0577 | 0.0581 | 0.0601 | 0.0578 | +0.0022 |
| | 5 | | 0.0595 | 0.0593 | 0.0615 | 0.0595 | +0.0021 |
| | 6 | | 0.0596 | 0.0594 | 0.0610 | 0.0592 | +0.0016 |
| | | | | | | | |
| Human RBC | 1 | 15 sec | 0.0553 | 0.0551 | 0.0572 | 0.0554 | +0.0019 |
| | 2 | | 0.0568 | 0.0571 | 0.0587 | 0.0570 | +0.0017 |
| | 3 | | 0.0550 | 0.0548 | 0.0571 | 0.0550 | +0.0022 |
| | | | | | | | |
| | 4 | 45 sec | 0.0548 | 0.0547 | 0.0565 | 0.0545 | +0.0018 |
| | 5 | | 0.0545 | 0.0548 | 0.0570 | 0.0547 | +0.0023 |
| | 6 | | 0.0558 | 0.0560 | 0.0575 | 0.0557 | +0.0017 |
| | | | | | | | |
| Beef Heart | 1 | 15 sec | 0.0600 | 0.0603 | 0.0632 | 0.0600 | +0.0031 |
| | 2 | | 0.0587 | 0.0587 | 0.0660 | 0.0586 | +0.0073 |
| | 3 | | 0.0601 | 0.0603 | 0.0642 | 0.0605 | +0.0039 |
| | | | | | | | |
| | 4 | 45 sec | 0.0670 | 0.0671 | 0.0713 | 0.0667 | +0.0044 |
| | 5 | | 0.0632 | 0.0636 | 0.0693 | 0.0633 | +0.0059 |
| | 6 | | 0.0669 | 0.0674 | 0.0725 | 0.0675 | +0.0052 |

| Porcine Heart | 1 | 15 sec | 0.0338 | 0.0340 | 0.0365 | 0.0340 | +0.0026 |
|---|---|---|---|---|---|---|---|
| | 2 | | 0.0321 | 0.0320 | 0.0357 | 0.0319 | +0.0037 |
| | 3 | | 0.0333 | 0.0333 | 0.0360 | 0.0333 | +0.0027 |
| | | | | | | | |
| | 4 | 45 sec | 0.0362 | 0.0363 | 0.0390 | 0.0362 | +0.0028 |
| | 5 | | 0.0312 | 0.0310 | 0.0382 | 0.0312 | +0.0071 |
| | 6 | | 0.0327 | 0.0329 | 0.0373 | 0.0331 | +0.0044 |
| | | | | | | | |
| Summary, all enzymes: $t_m$ = 15 sec, $\tau$ = 30 sec ||||||||

Table 3. Lactate – Chicken LDH Reaction $t_m$, $\tau$ Parameters

| | | ΔA340 / 12-sec enzyme reaction | | | | |
|---|---|---|---|---|---|---|
| **Run** | **t*** | **0 sec** | **t*-1 sec** | **t* sec** | **t*+1 sec** | **Δ(ΔA)** |
| 1 | 5 sec | 0.0628 | 0.0630 | 0.0649 | 0.0630 | +0.0020 |
| 2 | 5 sec | 0.0621 | 0.0620 | 0.0639 | 0.0622 | +0.0018 |
| 3 | 5 sec | 0.0636 | 0.0638 | 0.0654 | 0.0636 | +0.0017 |
| | | | | | | |
| 4 | 20 sec | 0.0658 | 0.0661 | 0.0679 | 0.0660 | +0.0019 |
| 5 | 20 sec | 0.0651 | 0.0648 | 0.0673 | 0.0650 | +0.0023 |
| 6 | 20 sec | 0.0639 | 0.0638 | 0.0656 | 0.0640 | +0.0017 |
| | | | | | | |
| Summary: $t_m$ = 5 sec, $\tau$ = 15 sec |||||||

Table 4. Comparison of Lactate Irradiation Activation for Chicken and Rabbit Muscle LDH

| | | ΔA340 / 12-sec enzyme reaction | | | | Δ(ΔA) | |
|---|---|---|---|---|---|---|---|
| **Run** | **t*** | **0 sec** | **t* sec** | **t* sec** | **0 sec** | **Chicken** | **Rabbit** |
| 1 | 5 sec | R 0.0662 | R 0.0664 | C 0.0730 | C 0.0703 | 0.0028 | 0.0002 |
| 2 | 5 sec | C 0.0703 | C 0.0732 | R 0.678 | R 0.0678 | 0.0029 | 0.0000 |
| | | | | | | | |
| 3 | 10 sec | R 0.0680 | R 0.0682 | C 0.0729 | C 0.0730 | 0.0001 | 0.0002 |
| 4 | 10 sec | C 0.0736 | C 0.0733 | R 0.0697 | R 0.0695 | 0.0003 | 0.0002 |
| | | | | | | | |
| 5 | 15 sec | R 0.0691 | R 0.0733 | C 0.0727 | C 0.0729 | 0.0002 | 0.0028 |
| 6 | 15 sec | C 0.0738 | C 0.0736 | R 0.0727 | R 0.0688 | 0.0002 | 0.0029 |
| | | | | | | | |
| 7 | 20 sec | R 0.0452 | R 0.0453 | C 0.0689 | C 0.0664 | 0.0025 | 0.0001 |
| 8 | 20 sec | C 0.0662 | C 0.0680 | R 0.0430 | R 0.0430 | 0.0018 | 0.0000 |
| | | | | | | | |
| 9 | 35 sec | R 0.0420 | R 0.0420 | C 0.0688 | C 0.0659 | 0.0029 | 0.0000 |
| 10 | 35 sec | C 0.0688 | C 0.0717 | R 0.0420 | R 0.0422 | 0.0029 | 0.0002 |
| | | | | | | | |
| Summary   Chicken: $t_m$=5 sec, **τ**=15 sec   Rabbit: $t_m$=15 sec (**τ**=30 sec) ||||||||

Table 5. Pyruvate – Mammalian LDH $t_m$, $\tau$ Parameters

|  |  |  | ΔA340 / 12-sec enzyme reaction | | | | |
|---|---|---|---|---|---|---|---|
| Source | Run | t* | 0 sec | t*-1 sec | t* sec | t*+1 sec | Δ(ΔA) |
| Rabbit Muscle Type V | 1 | 5 sec | 0.756 | 0.758 | 0.774 | 0.760 | +0.016 |
|  | 2 |  | 0.760 | 0.761 | 0.781 | 0.763 | +0.020 |
|  | 3 |  | 0.755 | 0.756 | 0.774 | 0.759 | +0.017 |
|  | 4 | 35 sec | 0.748 | 0.745 | 0.762 | 0.747 | +0.015 |
|  | 5 |  | 0.744 | 0.747 | 0.766 | 0.748 | +0.020 |
|  | 6 |  | 0.737 | 0.737 | 0.752 | 0.734 | +0.016 |
| Human RBC | 1 | 5 sec | 0.743 | 0.746 | 0.765 | 0.745 | +0.020 |
|  | 2 |  | 0.751 | 0.752 | 0.767 | 0.750 | +0.016 |
|  | 3 |  | 0.748 | 0.752 | 0.770 | 0.750 | +0.020 |
|  | 4 | 35 sec | 0.764 | 0.770 | 0.786 | 0.770 | +0.018 |
|  | 5 |  | 0.768 | 0.770 | 0.786 | 0.770 | +0.017 |
|  | 6 |  | 0.750 | 0.750 | 0.770 | 0.751 | +0.020 |
| Summary: $t_m$ = 5 sec, $\tau$ = 30 sec | | | | | | | |

Table 6.  Pyruvate – Chicken LDH Reaction $t_m$ Parameter

|  |  | ΔA340 / 12-sec enzyme reaction | | | |  |
|---|---|---|---|---|---|---|
| Run | t | 0 sec | t sec | t sec | 0 sec | Δ(ΔA) |
| 1 | 5 sec | 0.479 | 0.480 | 0.479 | 0.479 | +0.001 |
| 2 | 10 sec | 0.473 | 0.475 | 0.473 | 0.475 | +0.000 |
| 3 | 15 sec | 0.468 | 0.479 | 0.481 | 0.470 | +0.021 |
| 4 | 20 sec | 0.530 | 0.530 | 0.530 | 0.527 | +0.002 |
| 5 | 25 sec | 0.543 | 0.545 | 0.548 | 0.545 | +0.003 |
| 6 | 30 sec | 0.541 | 0.544 | 0.543 | 0.545 | +0.001 |
| Summary: $t_m$ = 15 sec | | | | | | |

Table 7.  Comparison of Pyruvate Irradiation Activation for Chicken and Rabbit Muscle LDH

|  |  | ΔA340 / 12-sec enzyme reaction | | | | Δ(ΔA) | Δ(ΔA) |
|---|---|---|---|---|---|---|---|
| Run | t* | 0 sec | t* sec | t* sec | 0 sec | Chicken | Rabbit |
| 1 | 5 sec | R 0.490 | R 0.508 | C 0.698 | C 0.702 | 0.004 | 0.018 |
| 2 |  | C 0.730 | C 0.730 | R 0.524 | R 0.504 | 0.000 | 0.020 |
| 3 | 10 sec | R 0.507 | R 0.507 | C 0.712 | C 0.713 | 0.001 | 0.000 |
| 4 |  | C 0.718 | C 0.719 | R 0.512 | R 0.513 | 0.001 | 0.001 |
| 5 | 15 sec | R 0.513 | R 0.516 | C 0.750 | C 0.723 | 0.027 | 0.003 |
| 6 |  | C 0.709 | C 0.731 | R 0.500 | R 0.500 | 0.022 | 0.000 |
| 7 | 20 sec | R 0.514 | R 0.514 | C 0.720 | C 0.718 | 0.002 | 0.000 |
| 8 |  | C 0.728 | C 0.725 | R 0.510 | R 0.508 | 0.003 | 0.002 |
| 9 | 25 sec | R 0.509 | R 0.509 | C 0.717 | C 0.719 | 0.002 | 0.000 |
| 10 |  | C 0.746 | C 0.743 | R 0.492 | R 0.495 | 0.003 | 0.003 |
| 11 | 30 sec | R 0.509 | R 0.506 | C 0.722 | C 0.725 | 0.003 | 0.003 |
| 12 | | C 0.723 | C 0.726 | R 0.505 | R 0.504 | 0.003 | 0.001 |
| | | | | | | | |
| 13 | 35 sec | R 0.550 | R 0.567 | C 0.734 | C 0.717 | 0.017 | 0.017 |
| 14 | | C 0.718 | C 0.736 | R 0.600 | R 0.582 | 0.018 | 0.018 |
| | | | | | | | |
| 15 | 55 sec | R 0.598 | R 0.598 | C 0.733 | C 0.715 | 0.018 | 0.000 |
| 16 | | C 0.715 | C 0.732 | R 0.584 | R 0.583 | 0.017 | 0.001 |
| | | | | | | | |
| Summary | Chicken: $t_m$ = 15 sec, $\tau$ = 20 sec    Rabbit: $t_m$ = 5 sec, $\tau$ = 30 sec | | | | | | |